\documentclass[aps,prd,preprint,nofootinbib]{revtex4-1}

\usepackage{amsmath,latexsym}
\usepackage{braket}
\usepackage{bbm}
\usepackage{color}
\usepackage{hyperref}
\usepackage{graphicx}

\newcommand{\BigO}[1]{\mathcal{O}\left(\frac1{c^{#1}}\right)}

\begin{document}

\title{\large Constraining noncommutative space-time from GW150914 \vspace{0.5cm}}

\author{Archil Kobakhidze}
\email[]{archil.kobakhidze@sydney.edu.au}
\author{Cyril Lagger}
\email[]{cyril.lagger@sydney.edu.au}
\author{Adrian Manning}
\email[]{adrian.manning@sydney.edu.au}
\affiliation{ARC Centre of Excellence for Particle Physics at the Terascale, School of Physics, The University of Sydney, NSW 2006, Australia \vspace{0.5cm}}

\begin{abstract}
The gravitational wave signal GW150914, recently detected by LIGO and Virgo collaborations,  is used to place a bound on the scale of quantum fuzziness of noncommutative space-time. We show that the leading noncommutative correction to the phase of the gravitational waves produced by a binary system appears at the second order of the post-Newtonian expansion. This correction is proportional to $\Lambda^2 \equiv |\theta^{0i}|^2/(l_P t_P)^2$, where $\theta^{\mu \nu}$ is the antisymmetric tensor of noncommutativity. To comply with GW150914 data, we find that $\sqrt{\Lambda} \lesssim 3.5$, namely at the order of the Planck scale. This is the most stringent bound on noncommutative scale, exceeding the previous constraints from particle physics processes by $\sim 15$ orders of magnitude.

\end{abstract}

\maketitle

\section{Introduction}

The LIGO/Virgo Collaboration recently announced the first direct detection of gravitational waves (GWs). The detected signal is referred to as GW150914, and was produced by the inspiral, merger and ringdown of a pair of black holes (BHs) \cite{Abbott:2016blz}. This observation provides the opportunity to test various fundamental physics scenarios as it is the first signal to probe gravity in such a strong and dynamical regime \cite{Yunes:2016jcc}. It is worth mentioning that no significant deviation from general relativity (GR) has been observed \cite{TheLIGOScientific:2016src}. In addition, many studies have already been conducted to investigate the implication of this signal, for example, to modifications of GR \cite{Yunes:2016jcc,Konoplya:2016pmh,Moffat:2016gkd,Vainio:2016qas}, the propagation of GWs \cite{Yunes:2016jcc,Blas:2016qmn,Collett:2016dey,Calabrese:2016bnu,Branchina:2016gad,Arzano:2016twc}, the search for dark matter under the form of primordial BHs \cite{Sasaki:2016jop,Bird:2016dcv,Clesse:2016vqa}, or the search for exotic compact objects \cite{Giudice:2016zpa}.

In this paper, we use the observation of GW150914 to constrain the scale of quantum space-time. The idea of 
considering quantized space-time by promoting space-time coordinates to noncommuting operators traces back to Heisenberg and was initially motivated to remove the ultraviolet divergences in quantum field theory. Since its first realization \cite{Snyder:1946qz}, the idea has gained renewed interest with the development of noncommutative geometry  \cite{Connes1985} and, especially, with the observation that quantized space-times represent the low-energy field-theoretic limit of string theory in the background of an antisymmetric B-field \cite{Ardalan:1998ce,Seiberg:1999vs} (see also reviews \cite{Douglas:2001ba,Szabo:2001kg}). In the latter approach, the space-time coordinate operators satisfy the canonical commutation relations
\begin{equation}
\label{equ:commutation_relation}
[\hat{x}^{\mu}, \hat{x}^{\nu}]=i \, \theta^{\mu \nu},
\end{equation}
where $\theta^{\mu \nu}$ is a real and constant antisymmetric tensor. Through $\theta^{\mu \nu}$ a new fundamental scale is introduced, which measures quantum fuzziness of space-time, similar to the Planck constant $\hbar$ that measures fuzziness of the phase space in the conventional quantum mechanics.

Various aspects of noncommutative field theories have been investigated in the past; see Refs. \cite{Douglas:2001ba, Szabo:2001kg} and references therein. Based on different treatments of noncommutative gauge symmetry, two different formulations of the noncommutative Standard Model of particle physics have been proposed in \cite{Chaichian:2001py, Chaichian:2004yw}  and \cite{Calmet:2001na, Aschieri:2002mc}. The limits on the noncommutative scale have been obtained from various particle physics processes, including low-energy precision measurements \cite{Mocioiu:2000ip, Chaichian:2000si} and processes involving Lorentz symmetry violation \cite{Carroll:2001ws, Calmet:2004dn}. In addition, inflationary observables can be used to constrain space-time noncommutativity \cite{K.:2014ssa, Calmet:2015fma}. Careful considerations \cite{Calmet:2004dn} show that the scale of noncommutativity is limited from these studies to be smaller than the inverse $\sim TeV$ scale. 

In addition, several versions of the noncommutative theory of gravitation have been suggested in Refs. \cite{Chamseddine:2000si,Aschieri:2005yw, Calmet:2005qm, Aschieri:2005zs,Kobakhidze:2006kb,Szabo:2006wx}. However, noncommutativity in all these formulations shows only  in the second order in the noncommutative scale \cite{Calmet:2006iz,Mukherjee:2006nd}, and thus the bounds from the purely gravitational sector are expected to be less restrictive. Bearing this in mind, we consider the effect of noncommutativity on GWs through the noncommutative corrections to the classical matter source and ignore noncommutative corrections to the gravity itself, which is highly model-dependent and presumably subdominant or even nonexistent as in string theory formulation.          

Under this assumption, it has already been shown by one of us \cite{Kobakhidze:2007jn} that the lowest-order noncommutative corrections to the matter source produce a second-order post-Newtonian modification of the Schwarzschild metric. We extend this analysis here to compute the noncommutative corrections to the waveform of the gravitational waves produced during the inspiraling phase of a BH binary system. We closely follow the post-Newtonian (PN) formalism \cite{Blanchet:2013haa} that allows analytical computation 
of equations of motion of a binary system and the associated radiation of gravitational waves within GR up to the order $\left(\frac{v}{c}\right)^7$ (or 3.5PN order\footnote{In this paper, we use the traditional post-Newtonian convention and say that a term of order $\left(\frac{1}{c}\right)^n \equiv\left(\frac{v}{c}\right)^n$ is a $\frac{n}{2}$PN term. }), where $c$ is the speed of light and $v$ a characteristic velocity of the system. Our calculations show that noncommutative effects in the energy-momentum tensor of the binary system imply a 2PN-order correction to the phase of the waveform produced by the BH pair. Comparing this analytical result to the numerical fitting of waveforms by LIGO and Virgo, we find the stringent limit on the time component of the noncommutative scale:
\begin{equation}
|\theta^{0i}| \lesssim 12 \cdot l_P t_P
\label{result}
\end{equation} 
where $l_P=\sqrt{\frac{\hbar G}{c^3}}\approx 1.6 \cdot 10^{-35}$ m and $t_P=\sqrt{\frac{\hbar G}{c^5}}\approx 5.4 \cdot 10^{-44}$ s are the Planck length and time. This bound is $\sim 15$ orders of magnitude smaller than the bound obtained from particle physics considerations.\footnote{We note in passing that $\theta^{0i}$ is considered to be vanishing within effective field theories, because of the apparent violation of unitarity (see, however, an alternative unitary formulation of noncommutative field theory with nonzero $\theta^{0i}$ in \cite{Balachandran:2004rq}). The potential violation of unitarity in the effective theory is not relevant for our calculations, and we simply assume that the issue is resolved in a full theory. Nonzero $\theta^{0i}$ also appears in unitary theories with lightlike noncommutativity, $\theta_{\mu\nu}\theta^{\mu\nu}=0$, which are known to have a consistent string theory completion \cite{Aharony:2000gz}.} 

The article is organized as follows. In Sec. \ref{sec2:NC_emt}, we derive the noncommutative corrections to the energy-momentum tensor describing two inspiraling BHs. In Sec. \ref{sec3:2PN_EOM}, we compute the equations of motion of this system including lowest-order, namely 2PN, noncommutative modifications. The energy flux radiated by the binary is then derived in Sec. \ref{sec4:flux}. Section \ref{sec5:bound} is devoted to the calculation of the phase of the waveform and to its comparison with the LIGO observation. We summarize our results and approximations in Sec. \ref{sec6:conclusion}. Throughout this work, we define the metric signature as $(-1,1,1,1)$. Latin indices range from $1$ to $3$ while greek indices range from $0$ to $3$.
 \section{\label{sec2:NC_emt}noncommutative corrections to the energy-momentum tensor}

A binary system is commonly approximated in GR by two point masses whose energy-momentum tensor is given by \cite{Blanchet:2000cw}
\begin{equation}
\label{equ:emt_GR_two_bodies}
T_{GR}^{\mu \nu}(\mathbf{x}, t) = m_1 \gamma_1(t) v_1^{\mu}(t) v_1^{\nu}(t) \delta^3(\mathbf{x}-\mathbf{y}_1(t)) + 1 \leftrightarrow 2
\end{equation}
with $m_i$ the masses; $\mathbf{y}_i(t)$ the positions; and $v^{\mu}_i(t)=\left( c, \frac{d \mathbf{y}_i(t)}{dt}\right)$ the  velocities of the two bodies $i=1,2$. The factor $\gamma_1$ is expressed through the metric $g_{\mu \nu}$ and its determinant $g$ as
\begin{equation}
\label{equ:gamma_factor_lorentz}
\gamma_1 = \frac1{\sqrt{g_1 (g_{\alpha \beta})_1 \frac{v_1^{\alpha} v_1^{\beta}}{c^2}}},
\end{equation}
and similarly for $\gamma_2$. In this expression, the metric and its determinant are evaluated at the location the body 1, namely, $(g_{\alpha \beta})_1 \equiv g_{\alpha \beta}(\mathbf{y}_1(t))$. Thus $\gamma_1$ only depends on time. However, the point-mass approximation implies that $g_1$ and $(g_{\alpha \beta})_1$ are divergent because of the delta functions in Eq. (\ref{equ:emt_GR_two_bodies}). This problem can be solved through the so-called Hadamard regularization whose application to the PN formalism is described in \cite{Blanchet:2000nu}. It is also worth mentioning that the energy-momentum tensor given by Eqs. (\ref{equ:emt_GR_two_bodies}) and (\ref{equ:gamma_factor_lorentz}) reproduces the correct GR equations of motion only up to 2.5PN order. For orders 3PN and higher, the spatial dependence needs to be accounted for in the metric \cite{Blanchet:2000ub}. Specifically, when evaluating the determinant of the metric in Eq. (\ref{equ:gamma_factor_lorentz}) we must use the field value instead of the location of the point mass, namely, $g_1$ has to be replaced by $g(t, \mathbf{x})$. Since the lowest-order noncommutative corrections will occur at 2PN order in the equations of motion, we can safely ignore this technicality in the present article.

In order to compute noncommutative corrections to the energy-momentum tensor (\ref{equ:emt_GR_two_bodies}), we follow the effective field theory formalism which has been used to compute quantum corrections \cite{BjerrumBohr:2002ks} and noncommutative corrections \cite{Kobakhidze:2007jn} to classical BHs. In this approach, the Schwarzschild BHs are sourced by a massive real scalar field $\phi$. To build a quantum field theory in noncommutative space-time, it is possible to work with the usual commuting coordinates $x^{\mu}$ instead of the operators $\hat{x}^{\mu}$ if we replace the product of two space-time-dependent functions by the following Moyal product \cite{Moyal:1949sk}:
\begin{equation}
\label{equ:moyal_product}
f(x)\star g(x) = f(x)g(x) + \sum_{n=1}^{+\infty} \left( \frac{i}{2}\right)^n \frac1{n!} \theta^{\alpha_1 \beta1} \cdots \theta^{\alpha_n \beta_n} \, \partial_{\alpha_1}\cdots \partial_{\alpha_n} f(x) \, \partial_{\beta_1} \cdots \partial_{\beta_n} g(x).
\end{equation}
The noncommutative energy-momentum tensor for a real scalar field $\phi$ (in natural units) can then be written as
\begin{equation}
\label{equ:emt_scalar_field_NC}
\begin{split}
T^{\mu \nu}_{NC}(x) &= \frac12 \left( \partial^{\mu} \phi \star \partial^{\nu} \phi + \partial^{\nu} \phi \star \partial^{\mu} \phi \right)-\frac12 \eta^{\mu \nu} \left( \partial_{\rho} \phi \star \partial^{\rho} \phi - m^2 \phi \star \phi \right) \\
& = \partial^{\mu} \phi \, \partial^{\nu} \phi -\frac12 \eta^{\mu \nu} \left( \partial_{\rho} \phi \, \partial^{\rho} \phi - m^2 \phi^2 \right) -\frac18 \theta^{\alpha_1 \beta_1} \theta^{\alpha_2 \beta_2} \Big(\partial_{\alpha_1} \partial_{\alpha_2} \partial^{\mu} \phi \partial_{\beta_1} \partial_{\beta_2} \partial^{\nu} \phi  \\
&  -\frac12 \eta^{\mu \nu} \partial_{\alpha_1} \partial_{\alpha_2} \partial_{\rho} \phi \partial_{\beta_1} \partial_{\beta_2} \partial^{\rho} \phi +\frac12 \eta^{\mu \nu} m^2 \partial_{\alpha_1} \partial_{\alpha_2} \phi \partial_{\beta_1} \partial_{\beta_2}  \phi \Big) +\cdots ~,
\end{split}
\end{equation}
where we only keep the lowest-order noncommutative corrections. Note that the first two terms correspond to the usual energy-momentum tensor of a massive scalar field. We then quantize the field in flat space-time as follows:
\begin{equation}
\label{equ:quantization_scalar_field}
\hat \phi(x)= \int \frac{d^3 \mathbf{k}}{(2 \pi)^3 \sqrt{2 \omega_{k}}} \left(\hat a(\mathbf{k}) e^{-i kx} + \hat a^{\dagger}(\mathbf{k})e^{ikx} \right),
\end{equation}
where
\begin{equation}
\label{equ:quantization_commutator}
\left[\hat a(\mathbf{k}), \hat a^{\dagger}(\mathbf{k}')\right]= (2 \pi)^3 \delta^3(\mathbf{k}-\mathbf{k}'), \quad \hat a(\mathbf{k}) \ket{0}=0, \quad \hat a^{\dagger}(\mathbf{k}) \ket{0}=  \ket{\mathbf{k}}.
\end{equation}
The expectation value of the energy-momentum tensor (\ref{equ:emt_scalar_field_NC}) between two arbitrary states $\ket{p_1}$ and $\ket{p_2}$ becomes at tree level:
\begin{equation}
\label{equ:matrix_element_emt}
\braket{p_2 | : \hat T_{NC}^{\mu \nu}(x): | p_1}= \frac{e^{- i\, \mathbf{q} \cdot \mathbf{x}}}{2 P^0} \left( 2 P^{\mu} P^{\nu} - \frac12 q^{\mu} q^{\nu} + \frac12 \eta^{\mu \nu} \mathbf{q}^2\right) \left(1 - \frac18 \left( \theta^{\alpha \beta} P_{\alpha} q_{\beta}\right)^2\right) ~,
\end{equation}
where we have defined $P=\frac12 ( p_1+p_2)$, and $q=p_1-p_2$, and we have chosen a frame in which $q^0=0$. Taking the Fourier transform of the previous formula, we obtain the position-space expression
\begin{equation}
\label{equ:fourier_transform_matrix_elem_emt}
\begin{split}
  & \int \frac{d^3 \mathbf{q}}{(2 \pi)^3} e^{i \, \mathbf{q} \cdot \mathbf{y}} \braket{p_2 | :  T_{NC}^{\mu \nu}(x): | p_1}  =   \\ &
 \frac{1}{2 P^0} \left(2P^{\mu} P^{\nu} + \frac12 \eta^{\mu m} \eta^{\nu n} \frac{\partial}{\partial x^m} \frac{\partial}{\partial x^n}- \frac12 \eta^{\mu \nu} \frac{\partial}{\partial x_i} \frac{\partial}{\partial x^i} \right)  \left(1 + \frac{\theta^{\alpha k} \theta^{\beta l} P_{\alpha} P_{\beta}}{8} \frac{\partial}{\partial x^k} \frac{\partial}{\partial x^l} \right)\, \delta^3(\mathbf{x}-\mathbf{y}) 
 \end{split}
\end{equation}
We now interpret the previous formula as the energy-momentum tensor of a pointlike particle of mass $m$ with momentum $P^{\mu}$ and position $\mathbf{y}(t)$, which we call $T_{NC,P}$. This expression can be further simplified once we restore the dimensions. Using the usual relativistic relation $P^{\mu}=m \gamma_L v^{\mu}$, where $\gamma_L$ is the Lorentz factor, we have
\begin{equation}
\label{equ:emt_NC_flat_general} 
\begin{split}
 T^{\mu \nu}_{NC,P }(\mathbf{x},t) =  & \, m \gamma_L v^{\mu} v^{\nu} \, \delta^3(\mathbf{x}-\mathbf{y}(t)) + \frac{m^3 \gamma_{L}^3 G^2}{8 c^4} v^{\mu} v^{\nu} \Theta^{kl} \partial_k \partial_{l} \, \delta^3(\mathbf{x}-\mathbf{y}(t)) \\
 + & \left( \eta^{\mu m} \eta^{\nu n} \partial_m \partial_{n} - \eta^{\mu \nu} \partial_i \partial^i \right) \left( \frac{\hbar^2}{4 m \gamma_L} +\frac{m \gamma_{L} \hbar^2 G^2}{32 c^4}  \Theta^{kl} \partial_k \partial_{l} \right) \, \delta^3(\mathbf{x}-\mathbf{y}(t)),
\end{split}
\end{equation}
where we have introduced\footnote{Note that the components $\theta^{0i}$ and $\theta^{ij}$ have different units.}
\begin{equation}
\label{equ:dimensonful_NC_tensor}
 \Theta^{kl} = \frac{\theta^{0k} \theta^{0l}}{l_P^2 t_P^2}  + 2 \frac{v_p}{c} \frac{\theta^{0k} \theta^{pl}}{l_P^3 t_P} + \frac{v_p v_q}{c^2} \frac{\theta^{kp} \theta^{lq}}{l_P^4}. 
\end{equation}
The second line of Eq. (\ref{equ:emt_NC_flat_general}) is proportional to $\hbar^2$, meaning that it will be negligible in front of the first two terms since we are considering astrophysical objects with  $m \gtrsim M_{\odot}$. On the other hand, Eq. (\ref{equ:dimensonful_NC_tensor}) tells us that in a post-Newtonian expansion  $ \Theta^{kl} = \frac{\theta^{0k} \theta^{0l}}{l_P^2 t_P^2} + \mathcal{O}\left(\frac{1}{c}\right)$. We will only keep this dominant term, since we are looking for the lowest-order noncommutative corrections to the waveform of the GWs produced by the binary system. From the two previous considerations, we can therefore simplify the energy-momentum tensor as follows:
\begin{equation}
\label{equ:emt_NC_flat_truncated}
 T^{\mu \nu}_{NC,P}(\mathbf{x},t) \approx   m \gamma_L v^{\mu} v^{\nu} \, \delta^3(\mathbf{x}-\mathbf{y}(t)) + \frac{m^3 \gamma_{L}^3 G^2}{8 c^4} v^{\mu} v^{\nu} \frac{\theta^{0k} \theta^{0l}}{l_P^2 t_P^2}  \partial_k \partial_{l} \, \delta^3(\mathbf{x}-\mathbf{y}(t)).
\end{equation}

We remind the reader that the previous expression has been derived in flat space-time. By identifying $\gamma_L$ with $\gamma_1$ defined by Eq. (\ref{equ:gamma_factor_lorentz}), the first term in Eq. (\ref{equ:emt_NC_flat_truncated}) reproduces exactly the GR energy-momentum tensor (\ref{equ:emt_GR_two_bodies}) for a single point mass. On the other hand, we do not need to generalize the second term to curved space-time since we are only interested in the lowest-order noncommutative corrections. By the same argument, we can also replace the Lorentz factor (in the second term) by its Newtonian value $\gamma_L = 1 + \BigO{2}$. This allows us to define the energy-momentum tensor of a binary system with its lowest-order noncommutative corrections as follows:
\begin{equation}
\label{equ:emt_starting_point}
\begin{split}
T^{\mu \nu}(\mathbf{x}, t) = & m_1 \gamma_1(t) v_1^{\mu}(t) v_1^{\nu}(t) \delta^3(\mathbf{x}-\mathbf{y}_1(t)) +  \frac{m_1^3  G^2 \Lambda^2}{8 c^4} v_1^{\mu}(t) v_1^{\nu}(t) \theta^k \theta^l \partial_k \partial_{l} \, \delta^3(\mathbf{x}-\mathbf{y}_1(t)) \\& +1 \leftrightarrow 2 ~,
\end{split}
\end{equation}
where we have simplified the notation by introducing $\Lambda \, \theta^i =\theta^{0i}/(l_P t_P)$, with $\theta^i$ representing the components of a three-dimensional unit vector $\boldsymbol{\theta}$, $\theta^i \theta^i=1$. In this way $\sqrt{\Lambda}$ corresponds to the time-component scale of noncommutativity relative to the Planck scale and will be the scale of noncommutativity we aim to constrain in this paper. 

 \section{\label{sec3:2PN_EOM} 2PN equations of motion}
\subsection{General orbit}
In order to infer the waveform of the GWs produced by the binary system, we require the equations of motion of the two bodies. Since we are neglecting noncommutative corrections to Einstein field equations (EFEs), we can invoke the covariant conservation of the energy-momentum tensor:
\begin{equation}
\label{equ:conservation_emt}
\nabla_{\nu} T^{\mu \nu}=0 \Rightarrow \partial_{\nu} \left(  \sqrt{-g} \, g_{\lambda \mu} T^{\mu \nu} \right)= \frac12 \sqrt{-g} \,  \partial_{\lambda} g_{\mu \nu} T^{\mu \nu}.
\end{equation}
Inserting the expression (\ref{equ:emt_starting_point}) for $T^{\mu \nu}$ in the previous equation gives the following relation for the first point-mass (considering only the spatial components):
\begin{equation}
\label{equ:EOM_P_and_F}
\frac{d P_1^i}{dt}=F_1^i,
\end{equation}
where, following the notation of \cite{Blanchet:2013haa}, the ``linear momentum density'' $\mathbf{P}_1$ and ``force density'' $\mathbf{F}_1$ satisfy
\begin{equation}
\label{equ:P_1st_expression}
P_1^i= \gamma_1 \, (g_{i \mu})_1 v_1^{\mu} + \frac{ m_1^2 G^2}{8 c^4} \Lambda^2 \theta^k \theta^l  \left( \partial_k \partial_{l} g_{i \mu}\right)_1 v_1^{\mu}
\end{equation}
\begin{equation}
\label{equ:F_1st_expression}
F_1^i= \frac12 \gamma_1 \, \left(\partial_{i} g_{\mu \nu}\right)_1 v_1^{\mu} v_1^{\nu} + \frac{ m_1^2 G^2}{16 c^4} \Lambda^2 \theta^k \theta^l    \left(\partial_k \partial_{l} \partial_{i} g_{\mu \nu}\right)_1 v_1^{\mu} v_1^{\nu}.
\end{equation}
The equations of motion of the second body are obtained by replacing the index 1 by 2 in the above formulas. Note that the expressions inside $(\cdots)_1$ have to be evaluated at the location of the body 1, $\mathbf{y}_1(t)$.

The previous equations describe the motion of the two point masses in the background of the metric $g_{\mu \nu}(\mathbf{x},t)$, which is itself generated by these two particles. The general form of this metric in the near zone of the system is given in the harmonic gauge in  Eq. (144) of \cite{Blanchet:2013haa}. At the order required for our analysis, it reads
\begin{equation}
\label{equ:metric_near_zone}
\begin{split}
g_{0 0}=-1 +\frac{2}{c^2} V + \BigO{4} \\
g_{0 i}=  \BigO{3}\ \\
g_{ij}= \delta_{ij} \left( 1 + \frac2{c^2} V \right) + \BigO{4}
\end{split}
\end{equation}
where $V$ is a retarded potential given by\footnote{Note that in the general PN formalism, some of the integrals in the definition of the retarded d'Alembertian may diverge at high PN order and require the use of a regularization technique; see details in \cite{Blanchet:2013haa}. This problem does not occur at the PN orders in which we are interested.}
\begin{equation}
\label{equ:retarded_potential_V}
V(\mathbf{x},t)=\Box_{\text{ret}}^{-1} \left[ - 4 \pi G \sigma\right] := G \sum_{k=0}^{+\infty} \frac{(-1)^k}{k!} \left( \frac{\partial}{c \, \partial t}\right)^k  \int d^3\mathbf{x}' |\mathbf{x}-\mathbf{x}'|^{k-1} \sigma(\mathbf{x}',t).
\end{equation}
This potential depends on the matter source through the quantity $\sigma = \frac{T^{00} + T^{ii}}{c^2}$. Using the expression (\ref{equ:emt_starting_point}) for $T^{\mu \nu}$ and keeping the leading noncommutating correction, we can rewrite it explicitly as
\begin{equation}
\label{equ:sigma_emt}
\sigma(\mathbf{x},t) = m_1 \gamma_1\left(1+\frac{v_1^2}{c^2}\right)\delta^3(\mathbf{x}-\mathbf{y}_1(t)) + \frac{m_1^3  G^2 \Lambda^2}{8 c^4}  \theta^k \theta^l \partial_k \partial_{l} \, \delta^3(\mathbf{x}-\mathbf{y}_1(t)) + 1 \leftrightarrow 2.
\end{equation}

It is worth observing that $V$ cannot be straightforwardly computed by inserting Eq. (\ref{equ:sigma_emt}) in Eq. (\ref{equ:retarded_potential_V}) since $\sigma$ depends on $V$ itself through $\gamma_1$ and $\gamma_2$. So $\sigma$ and $V$ are usually computed iteratively in each PN order. Fortunately, in our case of interest the lowest-order noncommutative correction to $V$ is simply computed by inserting the second term of Eq. (\ref{equ:sigma_emt}) into the $k=0$ term of the series (\ref{equ:retarded_potential_V}). In other words, we have
\begin{equation}
\label{equ:potential_V_2pn}
V(\mathbf{x},t)= V^{2PN}_{GR}(\mathbf{x},t)+ \frac{3 m_1^3 G^3 \Lambda^2}{8 c^4 r_1^3} \theta^k \theta^l  \hat{n}_{1kl} +\BigO{5} + 1 \leftrightarrow 2
\end{equation}
where $V^{2PN}_{GR}$ is the GR expression for the potential $V$ up to 2PN order, which can be explicitly found in Eq. (B1a) of \cite{Blanchet:1998vx}. We have also defined $r_1=|\mathbf{x}-\mathbf{y_1}|$, $\mathbf{n}_1=(\mathbf{x}-\mathbf{y}_1)/r_1$, and the symmetric trace free quantity $\hat{n}_{1kl}=n_{1k}n_{1l}-\delta_{kl}/3$.

We can now compute the 2PN expression of the ``linear momentum densities'' $\mathbf{P}_1$, $\mathbf{P}_2$ and ``force densities'' $\mathbf{F}_1$, $\mathbf{F}_2$. The first step is to introduce the metric (\ref{equ:metric_near_zone}) into Eqs. (\ref{equ:P_1st_expression})-(\ref{equ:F_1st_expression}):
\begin{equation}
\label{equ:P_2nd_expression}
P_1^i=  v_1^i  + \frac1{c^2} \left(P_1^{1PN}\right)^i + \frac1{c^4} \left(P_1^{2PN}\right)^i + \BigO{5}
\end{equation}
\begin{equation}
\label{equ:F_2nd_expression}
F_1^i= \left(\partial_i V\right)_1  + \frac1{c^2} \left(F_1^{1PN}\right)^i + \frac1{c^4} \left(F_1^{2PN}\right)^i +\frac{ m_1^2 G^2\Lambda^2}{8 c^4} \theta^k \theta^l \left(\partial_k \partial_{l} \partial_{i} V\right)_1 +  \BigO{5}
\end{equation}
and similarly for $\mathbf{P}_2$ and $\mathbf{F}_2$. The terms $P^{xPN}$ and $F^{xPN}$ represent some 1PN and 2PN expressions involving the retarded potential $V$ and some other higher-order retarded potentials [see Eqs. (146)-(147) and Eq. (152) in \cite{Blanchet:2013haa}]. The point is that noncommutative corrections to these terms appear above 2PN order in the final expression for $\mathbf{P}$ and $\mathbf{F}$ and are irrelevant in this study. Similarly, we observed that noncommutative corrections to $\mathbf{P}$ start at 3PN order and thus can be neglected, explaining their absence in Eq. (\ref{equ:P_2nd_expression}).

Thus only $\mathbf{F}$ admits 2PN-order corrections, which originate from both the first and fourth terms of Eq. (\ref{equ:F_2nd_expression}). Consider first $\left(\partial_i V\right)_1$ in which we replace $V$ by its expression (\ref{equ:potential_V_2pn}). Considering only the noncommutative term, we have
\begin{equation}
\label{equ:div_V_NC}
\left(\partial_i V_{NC}\right)_1=\left(- \frac{ 15 m_1^3 G^3 \Lambda^2}{8 c^4 r_1^4} \theta^k \theta^l \hat{n}_{1ikl} + 1 \leftrightarrow 2 \right)_1 = - \frac{ 15 m_2^3 G^3 \Lambda^2}{8 c^4 r^4} \theta^k \theta^l \hat{n}_{ikl}
\end{equation}
where $r=|\mathbf{y}_1-\mathbf{y}_2|$, $\mathbf{n}=(\mathbf{y}_1-\mathbf{y}_2)/r$, and $\hat{n}_{ikl}=n_{i}n_{k}n_{l}-\frac15 \left(n_{i}\delta_{lk}+n_{k}\delta_{il}+n_{l}\delta_{ik} \right)$.  At the second equality, we have used the Hadamard regularization described in \cite{Blanchet:2000nu}. Indeed, the first term in brackets is divergent when evaluated at the location of particle 1 since $r_1(\mathbf{y}_1(t))=0$. The second 2PN noncommutative correction to $\mathbf{F}_1$ comes from the fourth term in Eq. (\ref{equ:F_2nd_expression}) with $V$ replaced by its Newtonian value $V= G m_1/r_1+G m_2/r_2 + \BigO{2}$. We find after regularization:
\begin{equation}
\label{equ:second_correction_F}
 \frac{ m_1^2 G^2 \Lambda^2}{8 c^4} \theta^k \theta^l \left(\partial_k \partial_{l} \partial_{i} V\right)_1 = -\frac{15  m_1^2 m_2 G^3 \Lambda^2}{8 c^4 r^4} \theta^k \theta^l \hat{n}_{ikl}.
\end{equation}
By adding the two contributions (\ref{equ:div_V_NC}) and (\ref{equ:second_correction_F}), we have the final 2PN correction to the ``force density'' $\mathbf{F}_1$ as follows:
\begin{equation}
\label{equ:F_3rd_expression}
\left(F_1^i\right)^{2PN}_{NC} = -\frac{15  m_2(m_1^2+m_2^2) G^3 \Lambda^2 }{8 c^4 r^4} \theta^k \theta^l \hat{n}_{ikl}
\end{equation}
and similarly for the second body.

We now have all the ingredients to compute the acceleration of the first body in harmonic coordinates from Eq. (\ref{equ:EOM_P_and_F}). The calculation is performed iteratively at each PN order. The Newtonian part of $d P_1^i/dt$ gives $d v_1^i/dt=a_1^i$, which is directly compared to the Newtonian part of $F_1^i$. Then when the higher-order terms of $P_1$ are derived (e.g., $dP_1^{1PN}/dt$), each explicit acceleration that appears is order-reduced by its previous lower-order expression. Since there are no 2PN noncommutative corrections to $P_{1}^i$, it is straightforward to see that the only modification to the 2PN-order acceleration directly comes from the term (\ref{equ:F_3rd_expression}), namely,
\begin{equation}
\label{equ:radial_acceleration_body1}
a_1^i = (a_1^i)^{2PN}_{GR} - \frac{15 m_2(m_1^2+m_2^2) G^3 \Lambda^2}{8c^4 r^4} \theta^k \theta^l\hat{n}_{ikl}  +\BigO{5}.
\end{equation}
The GR acceleration $(a_1^i)^{2PN}_{GR}$ is given explicitly in Eq. (203) of \cite{Blanchet:2013haa} and has been computed iteratively following the procedure above. The acceleration of the second body is obtained by replacing the index 1 by 2 in the previous expression.

\subsection{Relative motion}
For the rest of the paper, we will consider only the relative motion of the two point masses. So in addition to $r$ and $\mathbf{n}$, we introduce the relative velocity $\mathbf{v}=\mathbf{v}_1-\mathbf{v}_2$ and acceleration $\mathbf{a}=\mathbf{a}_1-\mathbf{a}_2$. It is also useful to define the quantities
\begin{equation}
\label{equ:define_masses}
\renewcommand{\arraystretch}{1.5}
\begin{array}{lcl}
M &=& m_1+m_2 \\
\mu &=&\frac{m_1 m_2}{M} \\
\nu &=& \frac{\mu}{M} = \frac{m_1 m_2}{M^2} \\
\end{array}
\end{equation}
referred to, respectively, as the total mass, the reduced mass, and the symmetric mass ratio. From Eq. (\ref{equ:radial_acceleration_body1}), we directly deduce the relative acceleration
\begin{equation}
\label{equ:radial_acceleration_general}
a_i = (a_i)^{2PN}_{GR} - \frac{15 M^3 (1-2\nu) G^3 \Lambda^2}{8c^4 r^4}\theta^k \theta^l \hat{n}_{ikl}  +\BigO{5}.
\end{equation}

It has been proved \cite{deAndrade:2000gf} that the 2PN-order\footnote{Actually this result has been proved to 3PN order as well.} equations of motion in GR can be derived from a generalized Lagrangian $L^{2PN}_{GR}[\mathbf{y}(t), \mathbf{v}(t), \mathbf{a}(t)]$, which also depends on the acceleration. This Lagrangian is invariant under the Poincar\'{e} group and admits 10 Noetherian conserved quantities, including a conserved energy. Here ``conserved'' has to be understood in the sense of the post-Newtonian expansion. For example, the time derivative of a 2PN conserved quantity is at least of order $\BigO{5}$. In the same way, we can easily generalize the GR Lagrangian to take into account 2PN noncommutative corrections and we see that
\begin{equation}
\label{equ:lagrangian_2pn}
L = L^{2PN}_{GR} + \frac{3  M^3 \mu (1-2 \nu)G^3 \Lambda^2}{8 c^4 r^3}\theta^k \theta^l \hat{n}_{kl}  + \BigO{5} 
\end{equation}
reproduces the equations of motion (\ref{equ:radial_acceleration_general}). Note that the noncommutative part of this Lagrangian is not Lorentz invariant. However it still admits the following conserved energy:
\begin{equation}
\label{equ:energy_2pn_general}
E = E^{2PN}_{GR} -\frac{3  M^3 \mu (1-2 \nu) G^3\Lambda^2}{8 c^4 r^3}\theta^k \theta^l \hat{n}_{kl}   + \BigO{5}
\end{equation}
where $E^{2PN}_{GR}$ is given in Eq. (205) of \cite{Blanchet:2013haa}. Indeed, a direct computation\footnote{The time derivative of the second term in Eq. (\ref{equ:energy_2pn_general}) is canceled by the Newtonian part of $dE^{2PN}_{GR}/dt$ in which the acceleration has to be replaced by Eq. (\ref{equ:radial_acceleration_general}). } shows that $dE/dt=\BigO{5}$. 

In order to have a better understanding of the effect of the noncommutative terms in the acceleration and the energy, we can use the following identities:
\begin{equation}
\label{equ:stf_term_1}
\theta^k \theta^l \hat{n}_{ikl} = n_i \left(\mathbf{n} \cdot \boldsymbol{\theta}\right)^2- \frac15 n_i - \frac25 \theta_i \left(\mathbf{n} \cdot \boldsymbol{\theta}\right),
\end{equation}
\begin{equation}
\label{equ:stf_term_2}
\theta^k \theta^l \hat{n}_{kl}=\left(\mathbf{n} \cdot \boldsymbol{\theta}\right)^2- \frac13.
\end{equation}
In this form, we can see that the constant vector $\boldsymbol{\theta}$ acts like a preferred direction and will influence the motion of the binary system. In particular, we expect the orbital plane of the two point masses to precess because of the term $\theta_i \left(\mathbf{n} \cdot \boldsymbol{\theta}\right)$. On the other hand, the motion drastically simplifies if the orbital plane is perpendicular to this preferred direction as all the terms with $\mathbf{n} \cdot \boldsymbol{\theta}$ vanish. We argue now that we can restrict our attention to this simpler case since we are only looking for a bound on the parameter $\sqrt{\Lambda}$ and not a precise value.

There is of course no reason for the binary system that produced the GW150914 signal to satisfy this property. However it is important to observe that there are no orbital configurations for which each of the two expressions (\ref{equ:stf_term_1}) and (\ref{equ:stf_term_2}) are constantly zero, since $\boldsymbol{\theta}$ is time independent and $\mathbf{n}$ varies with time. In other words, the contributions $- \frac15 n_i$ in the acceleration and $-\frac13$ in the energy cannot be entirely canceled, they will only be modulated by the angular-dependent terms. Consequently, we expect the noncommutative corrections to the GW waveform to be of the same order of magnitude with or without these terms. Hence we will use the following expressions for the relative acceleration and energy of the binary:
\begin{equation}
\label{equ:radial_acceleration_simple}
a_i = (a_i)^{2PN}_{GR} + \frac{3 M^3 (1-2\nu) G^3 \Lambda^2}{8c^4 r^4} n_i  +\BigO{5}
\end{equation}

\begin{equation}
\label{equ:energy_2pn_simple}
E = E^{2PN}_{GR} +\frac{ M^3 \mu (1-2 \nu) G^3\Lambda^2}{8 c^4 r^3} + \BigO{5}.
\end{equation}

\subsection{Quasicircular orbit}
The previous equations further simplify if we assume that the two objects are in quasicircular orbit. This assumption is well justified since it has been shown in GR that the orbit of a binary system tends to circularize under the emissions of GWs \cite{Peters:1963ux,Peters:1964zz}. This is particularly true at the time when the GWs of the system enter the sensitivity band of the LIGO detector. It is important to note that this result has been directly derived from Einstein's quadrupole formula, which describes radiation of GWs at the lowest (Newtonian) order of the PN expansion. As we will explain in the next section, noncommutative corrections to the radiation formula appear at 2PN order, meaning that they are subdominant compared to the circularization effect. Therefore, even in noncommutative space-time, we expect to observe binaries with negligible eccentricity. Of course, precession of the orbital plane could still occur due to the angular-dependent terms in Eq. (\ref{equ:stf_term_1}). However, we shall not consider these terms for the reasons explained above. 

With these approximations we can assume that $r$ is constant, apart from the gradual inspiraling that will ultimately cause the two bodies to merge. Since this effect appears at 2.5PN order in the equations of motion \cite{Blanchet:1998vx}, we actually have $\dot{r}=\BigO{5}$. This condition greatly simplifies the GR part of Eqs. (\ref{equ:radial_acceleration_simple}) and (\ref{equ:energy_2pn_simple}). Indeed the relative acceleration in terms of the relative position $\mathbf{y}(t)=\mathbf{y}_1(t)-\mathbf{y}_2(t)$  reduces to
\begin{equation}
\label{equ:acceleration_circular}
\mathbf{a}_{circ}= - \Omega^2 \mathbf{y} + \BigO{5}.
\end{equation}
The angular frequency $\Omega$ is given by
\begin{equation}
\label{equ:omega_squared}
\Omega^2 = \frac{G M}{r^3} \left[ 1 + (-3 + \nu) \gamma + \left(6 + \frac{41}{4} \nu + \nu^2-\frac38 (1-2 \nu) \Lambda^2 \right) \gamma^2 \right] + \BigO{5}~,
\end{equation}
where we have introduced the following post-Newtonian parameter:
\begin{equation}
\label{equ:gamma_pn}
\gamma= \frac{G M}{c^2 r} = \BigO{2}.
\end{equation}
Note that the noncommutative term, proportional to $\Lambda^2$, comes directly from the second part of Eq. (\ref{equ:radial_acceleration_simple}), while all other terms come from the standard GR angular frequency $\Omega^2_{GR}$ given in \cite{Blanchet:2013haa} Eq. (228). In order to write the energy (\ref{equ:energy_2pn_simple}) of the two particles in circular orbit, we note that the norm $v$ of the relative velocity can be expressed as $v^2=r^2 \Omega^2 + \BigO{10}$. This implies in particular that the energy will contain two 2PN-order noncommutative corrections: one obvious contribution from the second term in Eq. (\ref{equ:energy_2pn_simple}) and another one from the Newtonian part of $E^{2PN}_{GR}$, once $v^2$ is expressed in terms of $\Omega^2$ given by (\ref{equ:omega_squared}). Adding these two contributions to the usual GR expression (229) in \cite{Blanchet:2013haa}, we find
\begin{equation}
\label{equ:energy_2pn_circular}
\begin{split}
 E_{circ}  = & - \frac{\mu c^2 \gamma}{2} \left[ 1 + \left( - \frac74 +\frac14 \nu\right) \gamma  + \left(-\frac78 + \frac{49}{8} \nu + \frac18 \nu^2 +\frac18 (1-2 \nu) \Lambda^2 \right) \gamma^2 \right] \\ & + \BigO{5}.
 \end{split}
\end{equation}

For later convenience, we want to express the energy as a function of the following frequency-related parameter:
\begin{equation}
\label{equ:x_frequency}
x = \left(\frac{G M \Omega}{c^3}\right)^{\frac23} =\BigO{2}.
\end{equation}
In order to achieve this relationship, we need to know how $\gamma$ depends on $\Omega$ (or $x$). We must therefore take the inverse of Eq. (\ref{equ:omega_squared}). We find
\begin{equation}
\label{equ:gamma_of_x}
\gamma = x \left[1 + \left(1-\frac13 \nu\right) x + \left(1-\frac{65}{12} \nu + \frac18 \Lambda^2 (1-2\nu)\right)x^2 \right] + \BigO{5}.
\end{equation}
By replacing $\gamma$ in Eq. (\ref{equ:energy_2pn_circular}), we finally obtain
\begin{equation}
\label{equ:energy_circular_of_x}
\begin{split}
E = &- \frac{\mu c^2 x}{2} \left[ 1+ \left( -\frac34 -\frac1{12} \nu \right) x + \left( -\frac{27}{8} + \frac{19}{8} \nu - \frac1{24} \nu^2 + \frac14 \Lambda^2 (1- 2 \nu) \right) x^2\right]  \\ &+ \BigO{5}.
\end{split}
\end{equation}
In Sec. \ref{sec5:bound}, we shall compare this expression to the energy radiated in GWs. 

 \section{\label{sec4:flux}Energy loss}
In this section we investigate the lowest-order noncommutative corrections to the energy radiated in GWs by the binary system. This energy loss is responsible for the secular decrease of the relative position $r$ between the two bodies and thus is a key ingredient to deduce the waveform of the emitted GWs. A first approach to compute this radiation would be to extend the calculation of the previous section to have the expression for the equations of motion and the energy of the system at higher PN orders. Indeed, it is known that the energy exhibits radiation-reaction terms starting from 2.5PN order \cite{Blanchet:2013haa}. However, the description of the inspiraling of the system at 2PN order would require knowing the equations of motion at 2PN order and the radiation terms at the 4.5PN level. This latter part is beyond the state-of-the-art knowledge in GR, since the energy is only known up to 3.5PN order. The second approach, which has been proven to be successful, is to identify this energy loss with the gravitational-wave flux $\mathcal{F}$ as seen by an observer far away from the source. In other words, we assume the following energy balance equation
\begin{equation}
\label{equ:balance_equation}
\frac{dE}{dt}=-\mathcal{F},
\end{equation}
where $E$ is given by Eq. (\ref{equ:energy_circular_of_x}), which is obtained from the equations of motion in the near zone. 

The expression for $\mathcal{F}$ is currently known to 3.5PN order in GR and to lowest-order corresponds to the well-known Einstein's quadrupole formula \cite{1918SPAW154E}. The general methodology (summarized in great detail in \cite{Blanchet:2013haa}) starts by solving the EFEs in the far zone of the source (with vanishing energy-momentum tensor) as a post-Minkowskian multipole expansion. This leads to an expression for $\mathcal{F}$ in terms of some radiative multipole moments [see Eq. (68) of \citep{Blanchet:2013haa}]. These radiative moments can then be expressed in terms of the source parameters through a well-justified matching strategy between the far zone and the near zone of the system. Since we have neglected noncommutative corrections to EFEs, this strategy is still valid in our case. The only modification to this procedure is to the source itself (\ref{equ:emt_starting_point}). From these considerations, the general form of the flux at 2PN order is
\begin{equation}
\label{equ:flux_general}
\mathcal{F}=\mathcal{F}_{inst}+\mathcal{F}_{tail},
\end{equation}
where $\mathcal{F}_{inst}$ is the so-called instantaneous flux, namely the flux produced only by the source multipole moments, and $\mathcal{F}_{tail}$ is composed of tail integrals coming from nonlinear multipole interactions between source and radiative moments (see Sec. 3.2 in \cite{Blanchet:2013haa}). Since the tail part of the flux starts at 1.5PN order, noncommutative corrections will appear above 2PN order. Hence we only need to compute corrections to the instantaneous part, which can be written as
\begin{equation}
\label{equ:flux_instant_general}
\mathcal{F}_{inst}= \frac{G}{c^5} \left[ \frac15  \frac{d^3 I_{ij}}{dt^3}\frac{d^3 I_{ij}}{dt^3} + \BigO{2} \right],
\end{equation}
where $I_{ij}$ is the mass quadrupole moment given as a function of the matter source $\sigma$ by Eq. (4.3) of \cite{Blanchet:1996wx}:
\begin{equation}
\label{equ:quadrupole_moment_general}
I_{ij} = \int d^3 \mathbf{x} \,\hat{x}_{ij} \sigma + \BigO{2}
\end{equation}
with $\hat{x}_{ij}= x_i x_j- \frac13 \delta_{ij} \mathbf{x}^2$. First, note that if $\sigma$ is replaced by its Newtonian expression, Eqs. (\ref{equ:flux_instant_general}) and (\ref{equ:quadrupole_moment_general}) reduce to Einstein's quadrupole formula. Second, it is worth emphasizing that higher-order terms, involving other source multipole moments, are required in both Eqs. (\ref{equ:flux_instant_general}) and (\ref{equ:quadrupole_moment_general}) to compute the flux $\mathcal{F}$ to 2PN order in GR. However, the previous equations are sufficient to calculate the lowest-order noncommutative corrections. 

We first investigate corrections to Eq. (\ref{equ:quadrupole_moment_general}) and perform the integral over the second term (noncommutative part) of $\sigma$ given by Eq. (\ref{equ:sigma_emt}):
\begin{equation}
\label{equ:integral_sigma_nc}
\begin{split}
\int d^3 \mathbf{x} \,\hat{x}_{ij} \sigma_{NC} &= \frac{m_1^3  G^2 \Lambda^2}{8 c^4}  \theta^k \theta^l \left. \left( \partial_k \partial_l \hat{x}_{ij}\right)\right|_{\mathbf{x}=\mathbf{y}_1(t)} + 1 \leftrightarrow 2 \\
&= \frac{(m_1^3+m_2^3)  G^2 \Lambda^2}{4 c^4} \left(\theta_i \theta_j- \frac13 \delta_{ij}\right).
\end{split}
\end{equation}
We directly observe that this term does not depend on time, meaning that this contribution will vanish in the instantaneous flux (\ref{equ:flux_instant_general}). Therefore the only nonvanishing 2PN noncommutative correction to $\mathcal{F}$ appears when we derive the Newtonian part of $I_{ij}$. Indeed, after two time derivations the expression will contain an acceleration that has to be replaced by the formula (\ref{equ:radial_acceleration_body1}), itself containing a 2PN correction. We now explicitly compute this correction assuming a circular orbit as we discussed in the previous section. In terms of the relative position $\mathbf{y}(t)$ of the two point masses, the quadrupole moment becomes
\begin{equation}
\label{equ:derivative_quadrupole_moment_general}
I_{ij} = \mu \left( y_i y_j- \frac13 \delta_{ij} r^2\right) + \BigO{2}.
\end{equation}
In taking the time derivative, the second term within the brackets vanishes since $\dot{r}=\BigO{5}$. The first term gives $d^3 y_i y_j/dt^3  = \dot{a}_i x_j + 3 a_i v_j+ (i\leftrightarrow j)$. Inserting the noncommutative term of the acceleration (\ref{equ:radial_acceleration_simple}) (and its time derivative), we find the following 2PN noncommutative correction:
\begin{equation}
\label{equ:third_derivation_xij}
\left.\frac{d^3}{dt^3} I_{ij}\right|_{NC} = \frac{3  \mu M^3 (1-2 \nu)G^3 \Lambda^2}{2 c^4 r^5} (y_i v_j+v_i y_j).
\end{equation}
Adding this term to the 2PN expression for the quadrupole mass moment in GR given in Eq. (C2a) of \cite{Blanchet:1995fg}, we obtain:
\begin{equation}
\label{equ:derivative_quadrupole_moment_2pn}
\begin{split}
\frac{d^3}{dt^3} I_{ij} =  & -\frac{ 8 G \nu M^2}{r^3} \left(\frac{y_i v_j + v_i y_j}{2}\right) \left[1 - \frac{\gamma}{42} \left(149-69 \nu \right) \right. \\
& \left. +\frac{\gamma^2}{1512} \left( 7043 - 7837 \nu + 3703 \nu^2 - 567 \Lambda^2(1-2 \nu) \right) \right] +\BigO{5}.
\end{split}
\end{equation}
The contribution from this quadrupole moment to the instantaneous flux is given by Eq. (\ref{equ:flux_instant_general}). Using the fact that in a circular orbit, $\mathbf{v}^2=\Omega^2 r^2 +\BigO{10}$ and $\mathbf{x}\cdot \mathbf{v}=\BigO{5}$, we have $(y_i v_j+v_i y_j)^2=2 r^4 \Omega^2+\BigO{5}$, where the angular frequency is given by Eq. (\ref{equ:omega_squared}). It follows straightforwardly that the lowest-order noncommutative correction to the flux is
\begin{equation}
\label{equ:flux_nc_correction}
\mathcal{F}_{NC}= \frac{32 c^5}{5G} \nu^2 \gamma^5 \left[ - \frac98 \Lambda^2 (1-2\nu)  \gamma^2 \right].
\end{equation}
The complete flux at 2PN order is computed by adding the previous term to the GR flux given in Eq. (4.16) of \cite{Blanchet:1995fg}. Note that the derivation of this latter expression requires higher-order multipole moments in Eq. (\ref{equ:flux_instant_general}) that we have not discussed since the noncommutative corrections to these terms will appear above 2PN in the final flux. As stated in Eq. (\ref{equ:flux_general}), tail effects are also important and will produce a 1.5PN contribution to $\mathcal{F}$. Taking all these contributions into account and after expressing $\gamma$ through $x$ thanks to (\ref{equ:gamma_of_x}), the final 2PN result including noncommutative corrections reads
\begin{equation}
\label{equ:flux_of_x_2pn}
\begin{split}
\mathcal{F} = & \frac{32 c^5}{5 G} \nu^2 x^5  \left[ 1 + \left(-\frac{1247}{336}-\frac{35}{12} \nu\right) x + 4 \pi x^{3/2} \right. \\
&  \left. + \left( - \frac{44711}{9072}+\frac{9271}{504} \nu + \frac{65}{18} \nu^2 - \frac{1}{2} \Lambda^2(1-2 \nu) \right) x^2 +\BigO{5}\right].
\end{split}
\end{equation}

 \section{\label{sec5:bound}Constraint on $\sqrt{\Lambda}$ from the orbital phase}

\subsection{Binary orbital phase}
We can now use the balance equation (\ref{equ:balance_equation}) to derive the secular decrease of the orbital radius $r$ and the rate of change of the orbital frequency $\Omega$. This will allow us to compute the evolution of the orbital phase of the binary system, which is a crucial parameter for data analysis. In order to do so, we first introduce the following dimensionless time variable,
\begin{equation}
\label{equ:dimensionless_time}
\Theta \equiv  \frac{ \nu c^3}{5 G M} (t_c-t) = \mathcal{O}\left( c^8\right),
\end{equation}
where $t_c$ represents the instant of coalescence of the two point masses. Obviously, the post-Newtonian formalism breaks down before the coalescence and is valid only during the inspiral of the binary. The description of the merger and the ringdown of the system typically requires the use of numerical methods to model accurately. Fortunately, as we shall see, the data obtained during the inspiral is sufficient to place a stringent bound on $\sqrt{\Lambda}$. 

In terms of $\Theta$, the energy balance equation becomes
\begin{equation}
\label{equ:ode_phase_general}
\frac{dE}{dx} \frac{dx}{d\Theta} = \frac{5 GM}{\nu c^3} \mathcal{F} ~,
\end{equation}
where $E(x)$ and $\mathcal{F}(x)$ are respectively given by Eqs. (\ref{equ:energy_circular_of_x}) and (\ref{equ:flux_of_x_2pn}). We recall that these quantities have been derived for a quasicircular orbit where the angular-dependent noncommutative terms in (\ref{equ:stf_term_1})-(\ref{equ:stf_term_2}) have been omitted. This expression  provides a differential equation for the frequency parameter $x(\Theta)$, which can be directly solved (in the PN expansion sense), giving
\begin{equation}
\label{equ:solution_ode_phase_2pn}
\begin{split}
x = & \frac14 \Theta^{-1/4} \left[1 + \left( \frac{743}{4032}+\frac{11}{48} \nu\right) \Theta^{-1/4} - \frac15 \pi \Theta^{-3/8} \right. \\ 
& \left. + \left( \frac{19583}{254016}+\frac{24401}{193536}\nu +\frac{31}{288} \nu^2 +\frac{10}{256} \Lambda^2 (1-2\nu) \right) \Theta^{-1/2} +\BigO{5} \right].
\end{split}
\end{equation}
This equation is nothing but the explicit temporal evolution of the angular frequency $\Omega$. It is then straightforward to find the orbital phase $\phi$ of the binary, defined as $d\phi/dt=\Omega$ or equivalently as
\begin{equation}
\label{equ:ode_orbital_phase}
\frac{d \phi}{d \Theta}= - \frac5{\nu} x^{3/2}.
\end{equation}
Integrating Eq. (\ref{equ:ode_orbital_phase}) with respect to $x$ [given in (\ref{equ:solution_ode_phase_2pn})] explicitly gives $\phi(\Theta)$. For data analysis purposes, it is more useful to express the phase in the frequency domain. So by inverting Eq. (\ref{equ:solution_ode_phase_2pn}), we can find $\Theta(x)$, allowing us to write the frequency-dependent phase evolution at 2PN precision:
\begin{equation}
\label{equ:orbital_phase_of_x}
\begin{split}
\phi = & -\frac{x^{-5/2}}{32 \nu} \left[ 1+ \left(\frac{3715}{1008} + \frac{55}{12} \nu \right) x - 10 \pi x^{3/2} \right. \\ 
& \left. + \left( \frac{15293365}{1016064}+\frac{27145}{1008} \nu + \frac{3085}{144} \nu^2 + \frac{25}{4} \Lambda^2 (1-2 \nu) \right) x^2 + \BigO{5} \right] ~,
\end{split}
\end{equation}
up to a constant of integration. We check that in the limit $\Lambda \rightarrow 0$, the two previous equations for $x$ and $\phi$ reduce to their GR expressions (316) and (318) in \cite{Blanchet:2013haa}. 

\subsection{Frequency-domain phasing template}
The previous results allow one to build analytical waveform templates that can be used to match to signals observed by the GW detectors. As mentioned previously, these analytical models can only describe the inspiral period of the coalescence since the post-Newtonian expansion breaks down due to the large velocities reached by the system in the later stages of the merger. For this reason, the later stages (merger and ringdown) are typically modeled using numerical relativity (NR). In the case of GW150914, the LIGO/Virgo Collaboration used two main waveform models combining both the PN expansion and NR: the effective-one-body (EOB) formalism \cite{Buonanno:1998gg,Damour:2009kr} and the IMRPhenom model \cite{Ajith:2009bn,Santamaria:2010yb,Khan:2015jqa}. 

For data analysis purposes, these templates are built in the frequency domain. Since the Fourier transform of the analytic time-domain templates cannot be performed analytically, it is usually computed under the stationary phase approximation (SPA). Starting from a GW signal with amplitude $A(t)$ and phase $\Phi(t)$ of the following form
\begin{equation}
\label{equ:waveform_cos}
h(t)=2 A(t) \cos \Phi(t),
\end{equation}
its Fourier transform in the SPA becomes \cite{Damour:2000zb}
\begin{equation}
\label{equ:waveform_frequency_domain}
\tilde{h}(f)=\frac{ \sqrt{2 \pi} A(t_f)}{\sqrt{\ddot{\Phi}(t_f)}} e^{i \psi(f)}, \quad \psi(f)=2 \pi f t_f - \pi/4 - \Phi(t_f).
\end{equation}
The parameter $t_f$ is the time when the GW frequency $d\Phi(t)/dt$ is equal to the Fourier frequency $f$. We can now relate the frequency-domain phase $\psi(f)$ to the orbital phase of the binary system. Indeed, it is known \cite{Blanchet:2013haa} that the GW frequency $d\Phi(t)/dt$ is twice the orbital frequency $\Omega$. Hence it implies that $t_f$ can be obtained by inverting Eq. (\ref{equ:solution_ode_phase_2pn}). Similarly, $\Phi(t_f)$ is obtained from $\phi(x)$ in Eq. (\ref{equ:orbital_phase_of_x}), recalling that $x$ is a frequency parameter. After a straightforward computation, we find the frequency-domain GW phase in terms of a PN expansion\footnote{The subscript $I$ is to emphasize that this formula is only valid during the inspiral stage.}:
\begin{equation}
\label{equ:phase_frequency_domain}
\psi_I(f)=2 \pi f t_c - \phi_c - \frac{\pi}{4} + \frac{3}{128 \nu} \sum_{j=0}^4 \varphi_j \left(\frac{\pi M G f}{c^3}\right)^{(j-5)/3},
\end{equation}
with $t_c$ and $\phi_c$ being the time and phase at coalescence. The coefficients in the previous expression are given by
\begin{equation}
\renewcommand{\arraystretch}{1.5}
\label{equ:coefficients_alpha}
\begin{array}{lcl}
\varphi_0 & = & 1 \\
\varphi_1 & = & 0 \\
\varphi_2 & = & \frac{3715}{756}+\frac{55}{9} \nu \\
\varphi_3 & = & -16 \pi \\
\varphi_4 & = & \frac{15293365}{508032}+\frac{27145}{504} \nu + \frac{3085}{72} \nu^2 + \frac{25}{2} \Lambda^2 (1-2\nu) ~.
\end{array}
\end{equation}
In standard space-time ($\Lambda=0$), we recover the coefficients used during the inspiral stage of the IMRPhenom template (see, e.g., \cite{Khan:2015jqa}). It should also be mentioned that in GR, the phase given in Eq. (\ref{equ:phase_frequency_domain}) is known up to 3.5PN order and therefore the previous list is completed with the coefficients $\varphi_5$ to $\varphi_7$, which we have ignored for our current purposes. We would expect noncommutative corrections to these terms as well, but their computation would be significantly more involved and are unnecessary to impose an initial constraint on $\sqrt{\Lambda}$. Also note that spin effects are usually included in the phase coefficients since it is known that spin-orbit and spin-spin terms appear from 1.5PN and 2PN order, respectively, in the equations of motion of a binary system (see Chap. 11 of \cite{Blanchet:2013haa} and references therein). But for the same reason as above, we do not consider these effects as well.

\subsection{GW150914 signal and constraint}
In \cite{TheLIGOScientific:2016src}, the LIGO/Virgo Collaboration used the GW150914 signal to test for deviations from GR. In their approach, they define a generalized IMR model (gIMR) whereby a phase deviation from standard GR, $\delta \varphi_j$, is introduced. This deviation is added to the IMRPhenom template by replacing the phase $\varphi_j$ with  $\varphi_j(1+\delta \varphi_j)$. The phase deviations $\left\{\delta \varphi_j \right\}$ are then allowed to vary (one at a time or all at once) in order to fit the theoretical template (including GR deviations) with the observation. In this way, a bound on each of these parameters can be inferred from a Bayesian analysis. The constraints derived from GW150914 are given in Table I of \cite{TheLIGOScientific:2016src}.

In order to find a robust constraint on the noncommutative scale $\sqrt{\Lambda}$, we would have to perform a similar analysis, namely, adding $\Lambda$ as a new parameter of the gIMR model and inferring it from a statistical analysis. However, an estimated bound can be computed using a significantly simpler method. From Eq. (\ref{equ:coefficients_alpha}), we define the fractional noncommutative deviation from GR as
\begin{equation}
\label{equ:delta_phi_4}
\delta \varphi_4^{NC} = \frac{\varphi_4^{NC}}{\varphi_4^{GR}}= \frac{1270080 \, (1- 2\nu)}{4353552 \nu^2 +5472432 \nu + 3058673} \Lambda^2.
\end{equation}
We then want to compare this correction to the value $\delta \varphi_4$ computed by LIGO/Virgo for GW150914. In order to do so we need the symmetric mass ratio $\nu$ of the binary system. However, it is important to realize that the masses of the BHs, $m_1=36.2^{+5.2}_{-3.8} M_{\odot}$ and $m_2=29.1^{+3.7}_{-4.4} M_{\odot}$ (in the source frame with $90\%$ credible regions), have been derived by LIGO \cite{Abbott:2016blz} from matched filtering based on GR templates. So if noncommutative corrections had been taken into account in those templates, we would expect slight deviations in the reported masses. Fortunately, this correction would have little significance on the constraint for $\sqrt{\Lambda}$. Indeed, by definition the symmetric mass ratio $\nu$ ranges between $0$ (test-mass limit) and $1/4$ (equal masses limit), which implies from Eq. (\ref{equ:delta_phi_4}) that
\begin{equation}
\label{equ:delta_phi_4_evaluated}
\delta \varphi_4^{NC} \in [1.35, 4.15] \cdot 10^{-1} \Lambda^2
\end{equation}
for any binary system. In other words, the indeterminacy in the masses has less than one order of magnitude impact on any constraints we can impose on $\Lambda$. Using the central values for $m_1$, $m_2$ given above, we have $\delta \varphi_4^{NC} =1.37 \cdot 10^{-1} \Lambda^2$.

In Table I of \cite{TheLIGOScientific:2016src}, LIGO computed that the deviation from GR of the fourth coefficient is given by $\delta \varphi_4 = -1.9^{+1.6}_{-1.7}$ when only $\delta \varphi_4$ is allowed to vary, and $\delta \varphi_4 = -1.9^{+19.3}_{-16.4}$ when all the coefficients can vary. Considering the worst case scenario and asking that $|\delta \varphi_4^{NC}| \lesssim |\delta \varphi_4|$, we derive the following estimated constraint:
\begin{equation}
\label{equ:bound_delta_phi_4}
|\delta \varphi_4^{NC}| \lesssim 20 \Rightarrow \sqrt{\Lambda} \lesssim 3.5 ~.
\end{equation}
Recalling that $\Lambda\equiv |\theta^{0i}|/(l_P t_P)$, the previous result means that the temporal part of the noncommutative tensor is constrained be around the Planck scale. 

Although we have not explicitly considered the spatial components $\theta^{ij} $ of the noncommutative tensor in this analysis, we would expect a similar constraint on them. From the energy-momentum tensor (\ref{equ:emt_NC_flat_general}) and Eq. (\ref{equ:dimensonful_NC_tensor}), it is clear that these terms would appear at 2.5PN and 3PN order in the equations of motion of the binary system (and in the energy flux). So we expect the coefficient $\varphi_6$ of the GW phase (\ref{equ:phase_frequency_domain}) to admit noncommutative terms proportional to $|\theta^{ij}|^2/l_P^4$. These terms would then be constrained as we just did, since LIGO computed \cite{TheLIGOScientific:2016src} that the deviation from GR of $\varphi_6$ is similar to $\delta \varphi_4$, namely, $\delta \varphi_6 = 1.2^{+16.8}_{-18.9}$.

 \section{\label{sec6:conclusion}Conclusion}

In this paper we have derived, to lowest-order, an analytic deviation from GR that would be present in the phase of gravitational radiation emitted from a binary black hole merger, should noncommutative space-time be manifest in nature. This deviation is dependent on the scale at which noncommutative space-time becomes prevalent. We show that (to lowest order) this phase deviation comes at 2PN order from a term proportional to $\Lambda^2 \equiv |\theta^{0i}|^2/(l_P t_P)^2$ and can be compared to the waveforms observed in the recent detection of gravitational waves from binary black hole mergers at LIGO, GW150914. By comparing the Bayesian analysis of allowed deviations from GR, which the LIGO and Virgo collaborations have completed using the GW150914 signal, we constrain $\Lambda$ up to the order of the Planck scale. 

In deriving this constraint, we made a number of well-justified approximations. First, we assumed that noncommutative effects contribute mainly to the energy-momentum tensor and ignored any corrections to the Einstein field equations. We expect these latter corrections to induce higher derivatives in the perturbed field equations and hence to be suppressed for low frequencies. Secondly, we have removed some angular-dependent terms in the noncommutative corrections to the equations of motion. As explained, these terms will only modulate the noncommutative corrections on the GW phase and hence have little effect on the overall bound that we placed on the scale of noncommutativity. Further, we assumed a circular orbit of the binary system, which is justified by the fact that emitted gravitational radiation removes angular momentum from a binary system and hence circularizes it. Finally, we used the masses of the binary black holes in the GW150914 estimated by the LIGO and Virgo collaborations, and we did not calculate these assuming noncommutative space-time. This is because deviations of the masses of the binaries play a very small role in constraining $\Lambda$ (as emphasized in Sec. \ref{sec5:bound}). 

Ultimately we find that if noncommutative space-time is realized its scale has to be of the order of the Planck scale in order to fit with the current measurements of gravitational waves from binary black hole mergers.

\section*{Acknowledgements}

We would like to thank Xavier Calmet for useful discussions. We are also grateful to anonymous referees for useful suggestions and especially for pointing out an omission in the earlier version of the manuscript. This work was partially supported by the Australian Research Council. A.K. was also supported in part by the Rustaveli National Science Foundation under the project No. DI/12/6-200/13.

\textit{Note added}.---Recently, the gravitational wave signal GW151226 detected by the LIGO detectors was announced \cite{PhysRevLett.116.241103}. The bound on space-time noncommutativity from this signal is the same order of magnitude as discussed in this paper.

\end{document}